# Natural analogs for geoengineering suggest modest radiative plant gains

Brendan Clark*[1], Ben Kravitz[2], Daniele Visioni[1], Manuel Lerdau[3]

[1]Department of Earth and Atmospheric Sciences, Cornell, University, Ithaca, NY

[2]Department of Earth and Atmospheric Sciences, Indiana University, Bloomington, IN

[3]Department of Environmental Sciences, University of Virginia, Charlottesville, VA

## Abstract

Stratospheric aerosol injection (SAI) to limit global warming overshoot would reduce total solar radiation while increasing the proportion of diffuse light reaching Earth's surface, creating competing effects on plant growth. Analog studies using volcanic eruptions show impacts ranging from significantly negative to strongly positive[1,2], and numerical models suggest radiative effects from SAI would range from neutral to slightly positive[3–5]. Here we show, using eddy covariance measurements at eight AmeriFlux sites (four maize and four forest), that radiative effects of smoke are positive across sites. This suggests that radiative effects from SAI would likely be modest but positive, consistent with numerical models. We attribute inconsistencies in the literature to limitations in the underlying assumptions and methods of previous analog studies; addressing these issues brings their results into agreement with our findings. More broadly, our results suggest that aerosols from wildfires, volcanoes, pollution, and SAI tend to moderately enhance photosynthesis.

## Main

With the world on track to exceed the global warming temperature targets set by the Paris Agreement, it is increasingly important to study potential ways to quickly limit warming overshoot[6]. Stratospheric aerosol injection (SAI) would mimic the cooling effects of major volcanic eruptions by scattering incoming solar radiation back to space, temporarily reducing surface temperatures at relatively low cost to mitigate the worst impacts of climate overshoot[7]. The consequent reduction in photosynthetically active radiation (PAR) can suppress photosynthesis. At the same time, SAI increases the fraction of surface radiation that is diffuse, allowing more light to penetrate deeper into the canopy and reach leaves that would otherwise be shaded under direct sunlight. This redistribution of light can enhance canopy-level photosynthesis[8,9].

Global numerical modeling studies have indicated that the counteracting effects of radiation from SAI would tend to have an overall net effect on crops and natural vegetation ranging from neutral to slightly positive[3–5]. However, studies using natural analogs for SAI to confirm or disprove such modeling studies have so far been limited. Gu et al.[2] found the radiative effects of the 1991 Mt. Pinatubo eruption to increase gross primary production (GPP) in the Harvard Forest by 23%. Mercado et al.[9] found that diffuse radiation from air pollution has increased the land carbon sink by 24% historically. However, Proctor et al.[1] used volcanic eruptions as an analog for SAI and concluded that its radiative effects could significantly reduce crop yields, potentially offsetting the benefits of lower temperatures.

Here we utilize eddy covariance measurements at four cropland and four forested AmeriFlux sites to understand how changes in radiation during wildfire smoke events impacts GPP, in an attempt to resolve these discrepancies. Wildfire smoke events, similar to the sulfate aerosols emitted during large volcanic eruptions and proposed for SAI, increase the proportion of diffuse radiation reaching the surface while simultaneously reducing total incoming solar radiation. They provide a more recent analog for SAI, offering additional observational data to better understand its potential radiative effects.

We show that the radiative effects of wildfire smoke events create positive effects on GPP for both cropland and forested ecosystems. Applying this relationship to a SAI scenario from climate model output suggests small, positive radiative effects on photosynthesis. We show that these results are consistent with previous studies examining the impacts of air pollution, and that even modest changes in GPP at the global scale could translate into substantial increases in the terrestrial carbon sink[9]. We also re-examine previous studies of the radiative impacts of volcanic eruptions on vegetation[1,2], identifying limitations in their underlying assumptions and methods. Correcting for these limitations brings their findings into closer agreement with our results and those of Mercado et al.[9].

Overall, these results suggest that potential radiative effects of SAI on photosynthesis in both managed and unmanaged ecosystems would be modest and positive, and that reconciled analog studies are broadly in agreement with numerical model simulations. Large wildfire events have been increasing in the U.S. over the past 20 years, and further study of how this smoke may impact plant growth will provide benefits beyond analogs of SAI (Figure 1).

### Wildfire smoke and sulfate impacts on photosynthesis

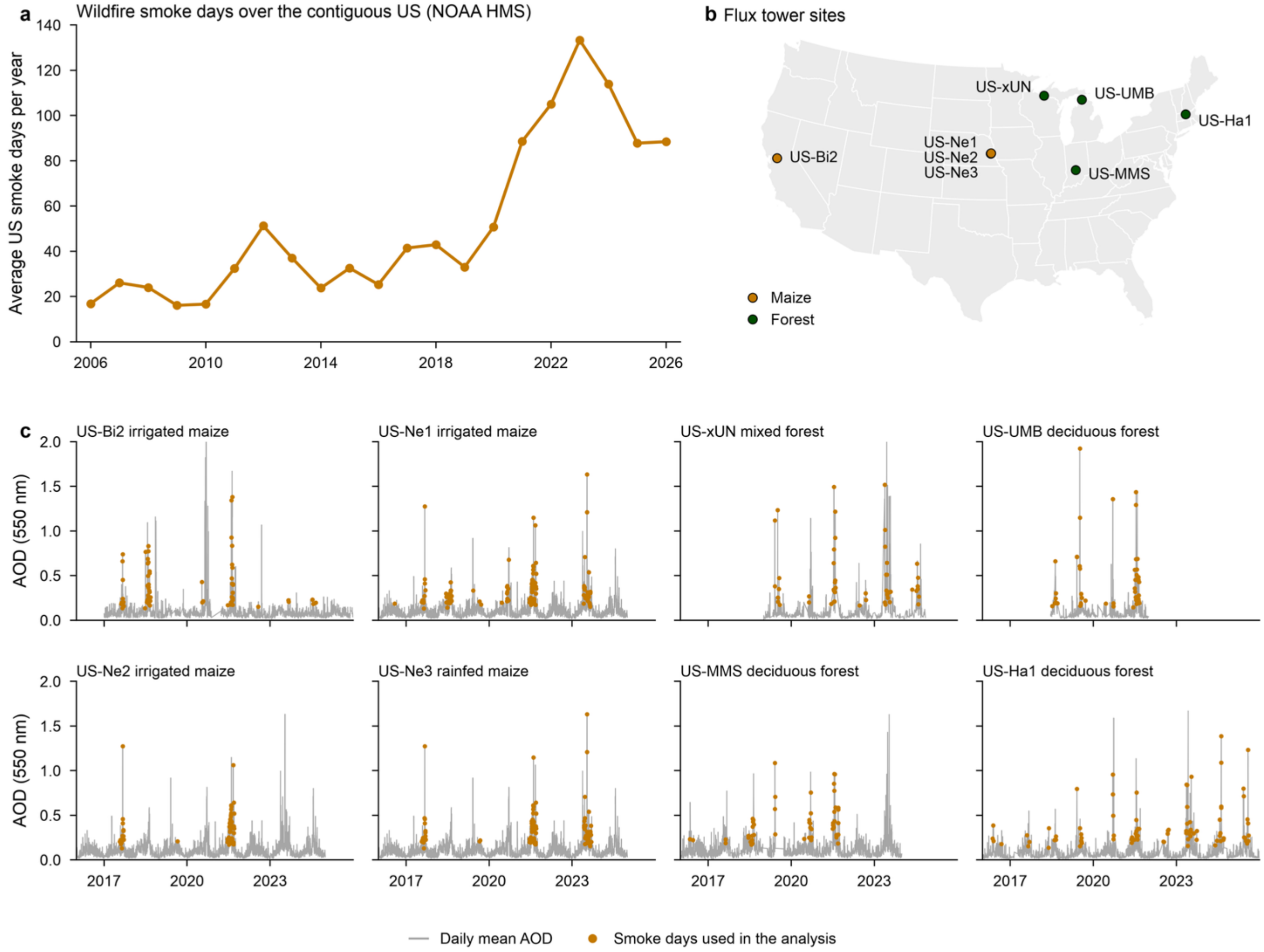

**Figure 1. Wildfire smoke over the contiguous US and at eight AmeriFlux sites.**

(a) The average number of smoke days over the contiguous U.S. as monitored by NOAA HMS. (b) A map showing locations of all AmeriFlux sites used in the analysis (c) The average daily 550 nm AOD at each AmeriFlux site over the period analyzed, including periods where smoke was identified.

Over each AmeriFlux site, we identify periods of clear-sky smoke using NOAA's Hazard Mapping System Fire and Smoke Product (HMS), and AERONET measurements of aerosol optical depth (AOD) near each site. These smoke events are then compared to non-smoke clear sky measurements with similar conditions. We estimate the smoke effects of changes in total PAR, diffuse fraction, temperature, soil moisture and vapor pressure deficit (VPD) on GPP using a site-specific ordinary least squares regression, fitted to growing-season hours under clear and partly cloudy skies. Including partly cloudy hours widens the range of light and diffuse conditions, which helps separate the effect of diffuse fraction from that of total light.

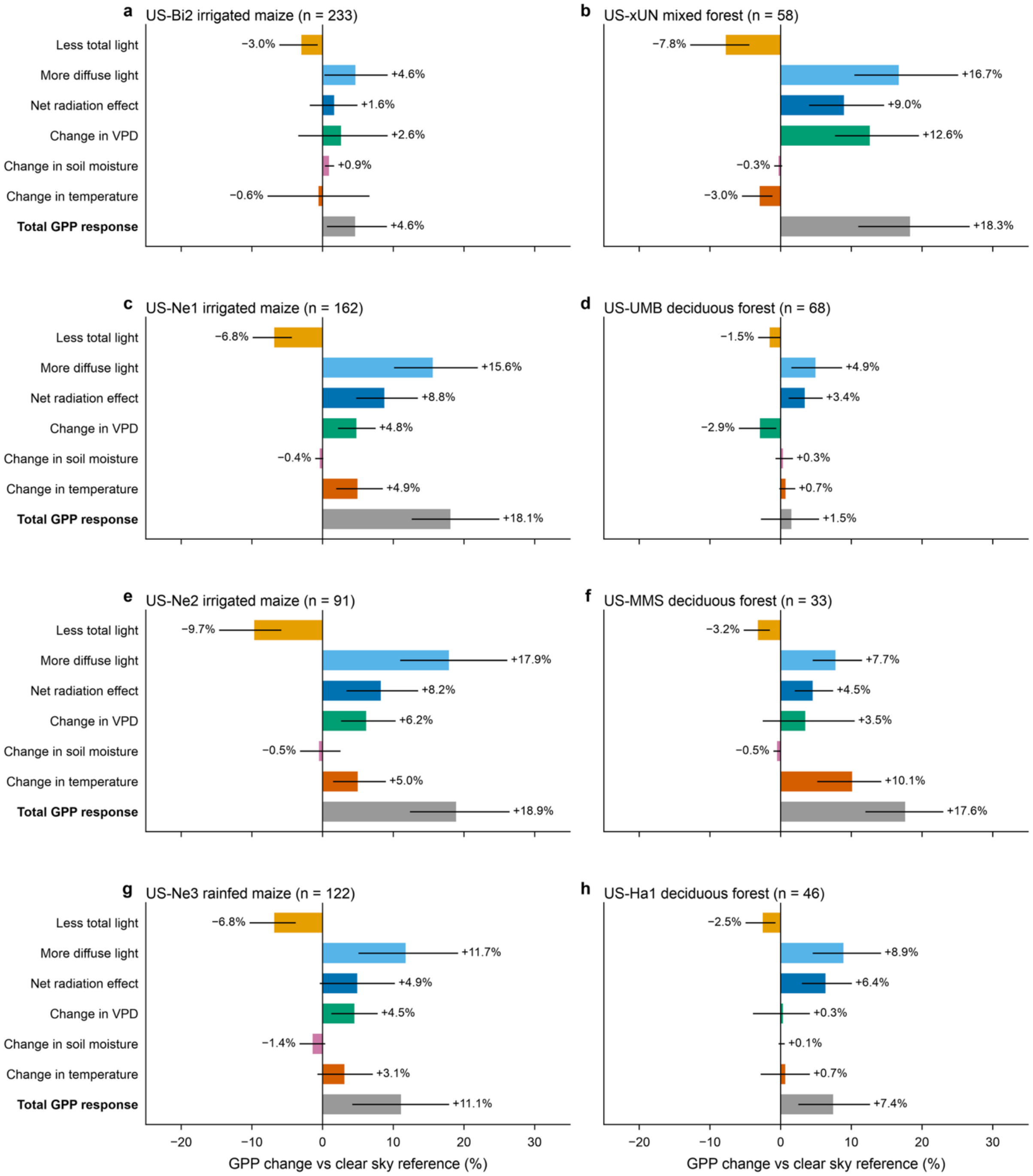

**Figure 2. Wildfire smoke increases GPP through radiation and weather at all eight sites.**

The individual and total effects of smoke on GPP at each AmeriFlux site. The total GPP response is the sum of the net radiation, VPD, soil moisture and temperature effects. Error bars are 95% confidence intervals from 1,000 bootstrap samples that resample whole days. n is the number of matched smoke time steps.

Across the eight AmeriFlux sites, wildfire smoke-induced radiation perturbations never decreased GPP, with seven of the eight sites exhibiting significantly positive GPP responses (Figure 2). Wildfire smoke also tends to lower air temperature and VPD by reducing the solar radiation reaching the surface, which increased GPP at many of the sites. However, we cannot separate these changes from those caused by the weather conditions that accompany smoke events. Soil moisture is not significantly affected by wildfire smoke at any sites. The total effect of smoke increases overall GPP by 4.6 to 18.9% for maize sites, and by 1.5 to 18.3% for forest sites (Figure 2).

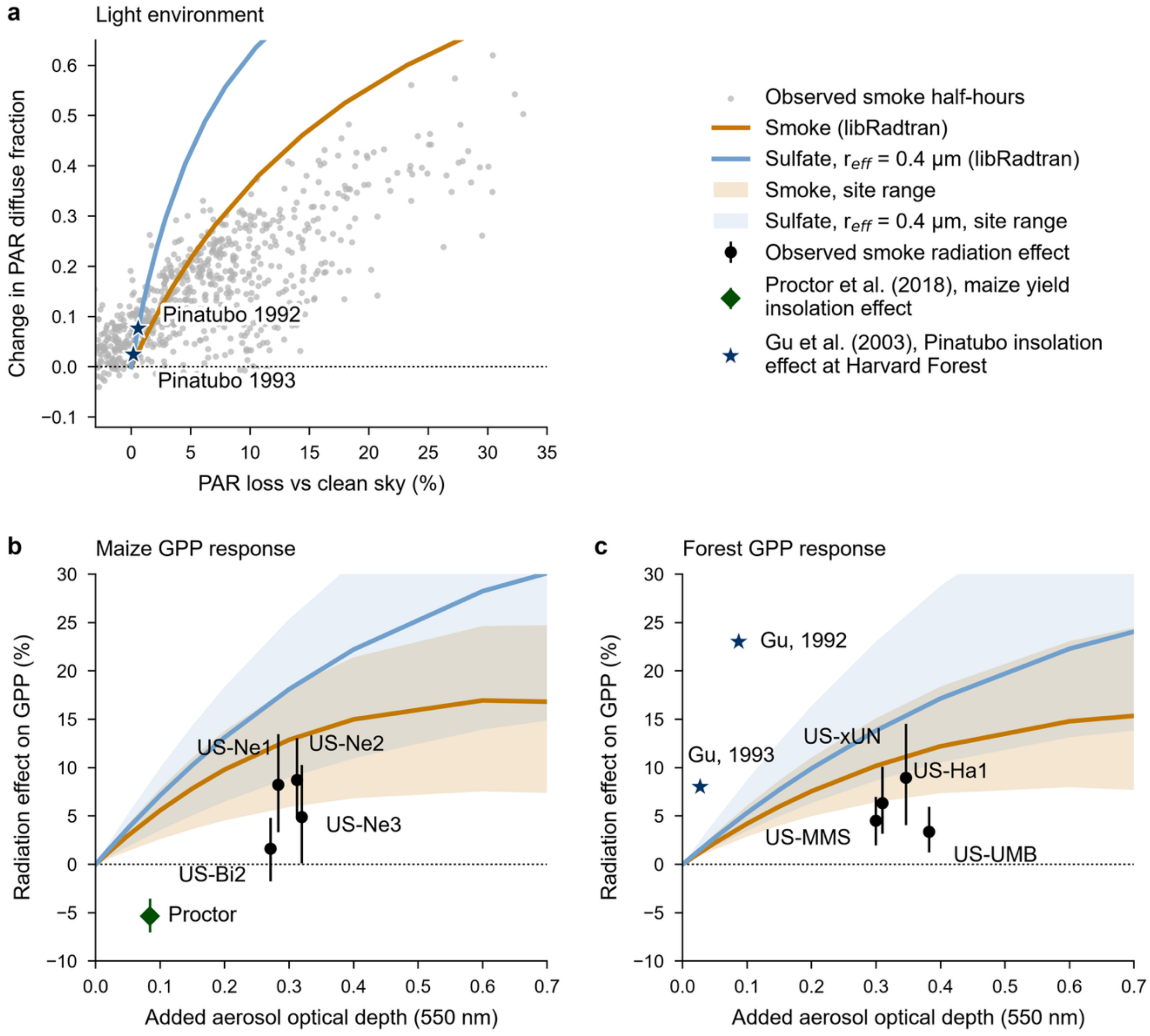

**Figure 3. Light changes and GPP responses for smoke and sulfate aerosol.**

(a) Change in PAR diffuse fraction versus PAR loss for observed smoke time steps (grey points), and for smoke and sulfate aerosol simulated with libRadtran (lines, see Methods for details). Stars indicate Pinatubo values based on Gu et al.[2]. (b, c) Radiation effect on maize (b) and forest (c) GPP of the libRadtran-simulated smoke and sulfate light changes, from the site regressions fitted to AmeriFlux observations. Blue and orange lines show the mean of the four sites and shading the range across sites. Black points are the observed smoke radiation effects with 95% confidence intervals (as in Fig. 2).

Figure 3 shows that libRadtran, a radiative transfer model[10] (see Methods for details), tends to overestimate the fraction of diffuse radiation from wildfire smoke. For a given AOD, the effects of SAI with sulfate would increase GPP more than smoke. Results from Proctor et al.[1] and Gu et al.[2] fall well outside of the range seen in any of the AmeriFlux sites. Proctor et al.[1] use country-level crop yields rather than site-level photosynthesis, so their results are not directly comparable. However, Gu et al.[2] report results from Harvard Forest, which is included in our analysis (US-Ha1).

## Comparison with previous volcanic eruption studies

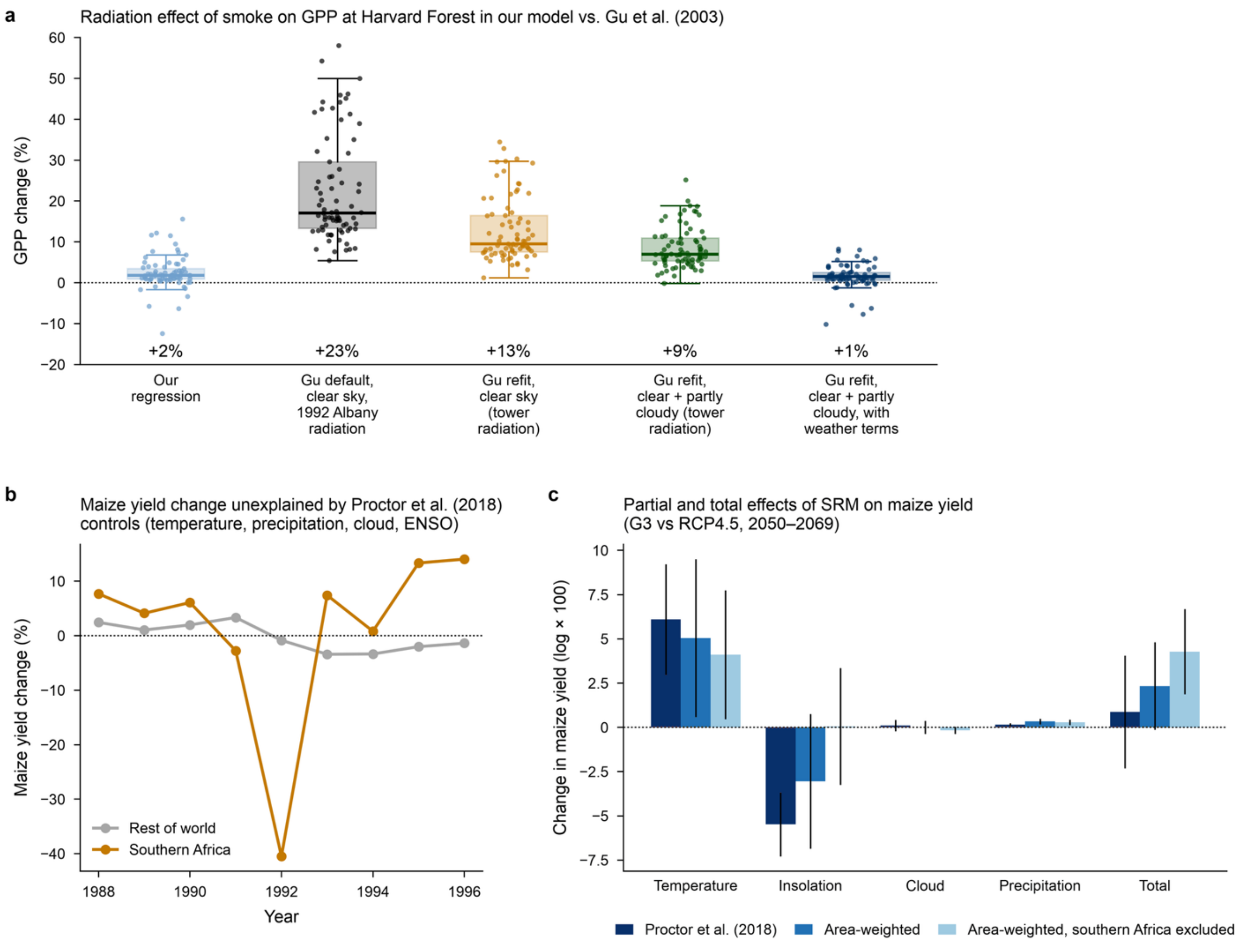

**Figure 4. Correcting earlier volcanic analog studies brings them into agreement with smoke observations.**

(a) Effects of wildfire smoke-induced radiation changes on GPP at Harvard Forest. The light blue boxplot (far left) shows our regression model, and the black boxplot shows the default model from Gu et al.[2], driven by 1992 radiation from Albany, NY. The yellow boxplot shows Gu et al.'s model refitted to all available clear-sky tower data from Harvard Forest. The green boxplot shows the model refitted to both clear-sky and partly cloudy data, and the dark blue boxplot also adds weather terms to the refitted model. Points are individual smoke days, and percentages give the overall effect. To allow a direct comparison with Gu et al.[2], periods of extreme heat and high atmospheric demand are not filtered out. (b) Maize yield changes not explained by Proctor et al.'s[1] controls (temperature, precipitation, cloud cover and El Niño Southern Oscillation), averaged over southern Africa (South Africa, Zimbabwe, Zambia, Malawi, Mozambique, Botswana, Namibia, Lesotho and Eswatini) and the rest of the world, 1988–1996. (c) Projected effects of solar radiation modification (G3 SAI scenario vs RCP4.5, 2050–2069; see Proctor et al.[1]) on maize yield through temperature, insolation, cloud cover and precipitation, and their total. Bars show Proctor et al.'s regression as published, the same regression with each country weighted by its maize harvested area, and the area-weighted regression with southern Africa excluded. Error bars are 95 % confidence intervals.

To reconcile these discrepancies, we investigate the methods of Gu et al. and Proctor et al. more thoroughly. Revisiting Gu et al.[2] reveals that their model substantially overestimates the photosynthetic benefits of diffuse radiation when applied to wildfire smoke at Harvard Forest. Using their published parameters, the model predicts a 23% increase in GPP, compared with approximately 2% from our regression (Fig. 4a). However, refitting the model using a longer observational record that includes both clear sky and partly cloudy conditions, while accounting for weather effects, reduces the estimated increase to just 1%, closely matching our results (Fig. 4a). This demonstrates that the discrepancy between their results and ours arises primarily from the limited data and fitting approach used in the original study, rather than differences in model structure. Relying on a single year of clear-sky observations can make the model unable to separate the responses to diffuse and direct light, because diffuse fraction varies little under clear skies, and leaves its parameters sensitive to the weather of that one year. Although Gu et al.[2] argue that their comparison is unaffected by weather because the same parameters are used to model photosynthesis under both the post-eruption and non-perturbed radiation conditions, their model is trained on data in which a diffuse-light benefit can coincide with favorable weather (e.g., lower VPD or temperature). Correcting these two limitations makes their results closely match ours, with a modest radiative benefit.

Revisiting Proctor et al.[1] reveals that their reported negative effects of radiation from volcanic aerosols on crop yields are sensitive to how regional data are weighted and to anomalous post-eruption climate conditions. Their regression is on country-level data and applies no area weighting, so the estimated global response is the average response of each country in their sample, regardless of how much of the crop it grows. Weighting each country by its maize harvested area in their regression model (Eq. 10, Methods) changes the estimated negative effect of radiation from volcanic eruptions on maize yield, from −5.4

% to −3.0 % (Figure 4c.), and the 95 % confidence interval of the weighted estimate includes zero. For the other crops (rice, soybean, and wheat), area weighting causes the insolation effect to be indistinct from zero (Extended Data Figure 2). Area-weighting the results also allows a better comparison to crop model simulations, as they typically operate under a fixed map of harvested area, and their regional and global yield changes are aggregated by weighting each grid cell by its crop area[11]. The remaining negative maize response is largely driven by the severe 1992 El Niño drought in southern Africa, which reduced yields by approximately 40% below the levels predicted by Proctor et al.'s controls (Fig. 4b). This may be due to Proctor et al.'s use of a single yield-vs-rainfall curve for every country in the analysis, or one global El Niño term which may not capture how hard El Niño hit southern Africa in 1992. Because this drought was unrelated to Pinatubo[12] but occurred in the year following the eruption, its effects were erroneously attributed to insolation. Excluding the nine affected countries further reduces the estimated effect, and combined with harvested-area weighting, brings the estimated insolation effect on maize yields to approximately zero (Fig. 4c). Thus, accounting for differences in crop production across countries and the effects of El Niño not captured by the original controls eliminates the previously reported negative yield response to volcanic aerosols.

## Radiative effects of stratospheric aerosol injection

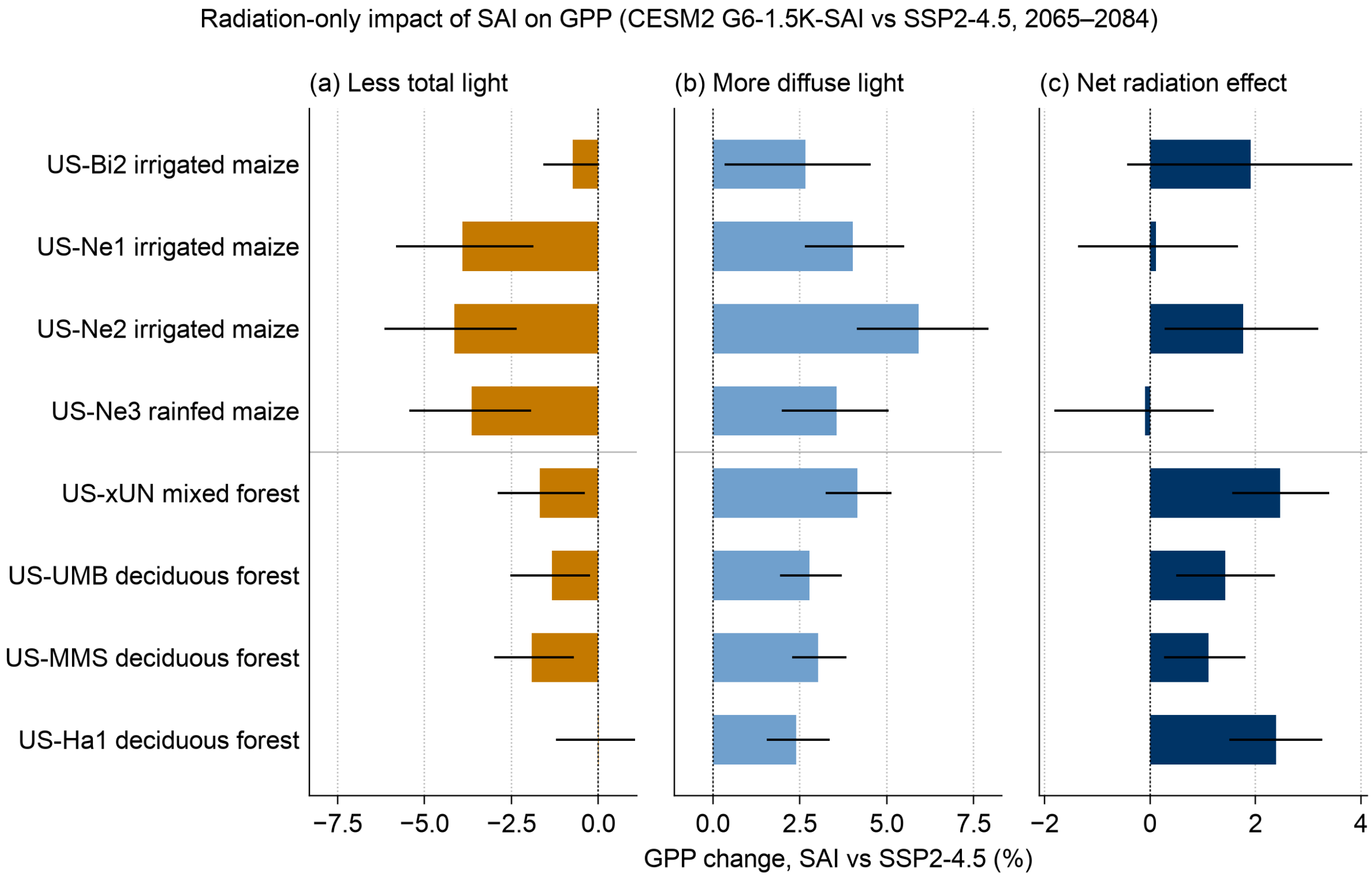

**Figure 5. Stratospheric aerosol injection would modestly increase GPP through radiation.**

The regression estimated impact of changes in total light, diffuse light, and the net effect under stratospheric aerosol injection relative to climate change as simulated by the Community Earth System Model[13].

Figure 5 shows the change in GPP at each of the eight sites under SAI, calculated by applying the CESM2-simulated changes in light and diffuse fraction (G6-1.5K-SAI vs. SSP2-4.5, 2065–2084) to each site's tower observations and running them through that site's regression (see Methods for details). At six sites, the benefit from enhanced diffuse radiation outweighs the reduction in GPP from less total radiation, with two sites showing the net effect to be zero. Overall, most sites show an increase to GPP under SAI of about 2% due to more favorable radiation (Figure 5). Two out of the four maize sites suggest no effect of radiation from SAI on GPP, consistent with the revised Proctor et al.[1] results (Figure 4c). While these increases seem modest, small changes in GPP (propagated over the entire globe) can lead to large increases in the land carbon sink because the land sink (NEE) is only a small fraction of two large fluxes, photosynthesis and ecosystem respiration[9]. Figure 6 shows that an increase in the land carbon sink of 24% due to air pollution was driven by a 1% increase in GPP. Normalized by change in diffuse fraction, the results of Mercado et al.[9] closely match ours (Figure 6b). This suggests that the modest GPP increase projected under SAI (about +2 %, Figure 5), if it held globally, could substantially increase the land carbon sink. However, this enhancement may weaken over time as respiration from the enlarged vegetation and soil carbon pools catches up with the increased photosynthetic uptake.

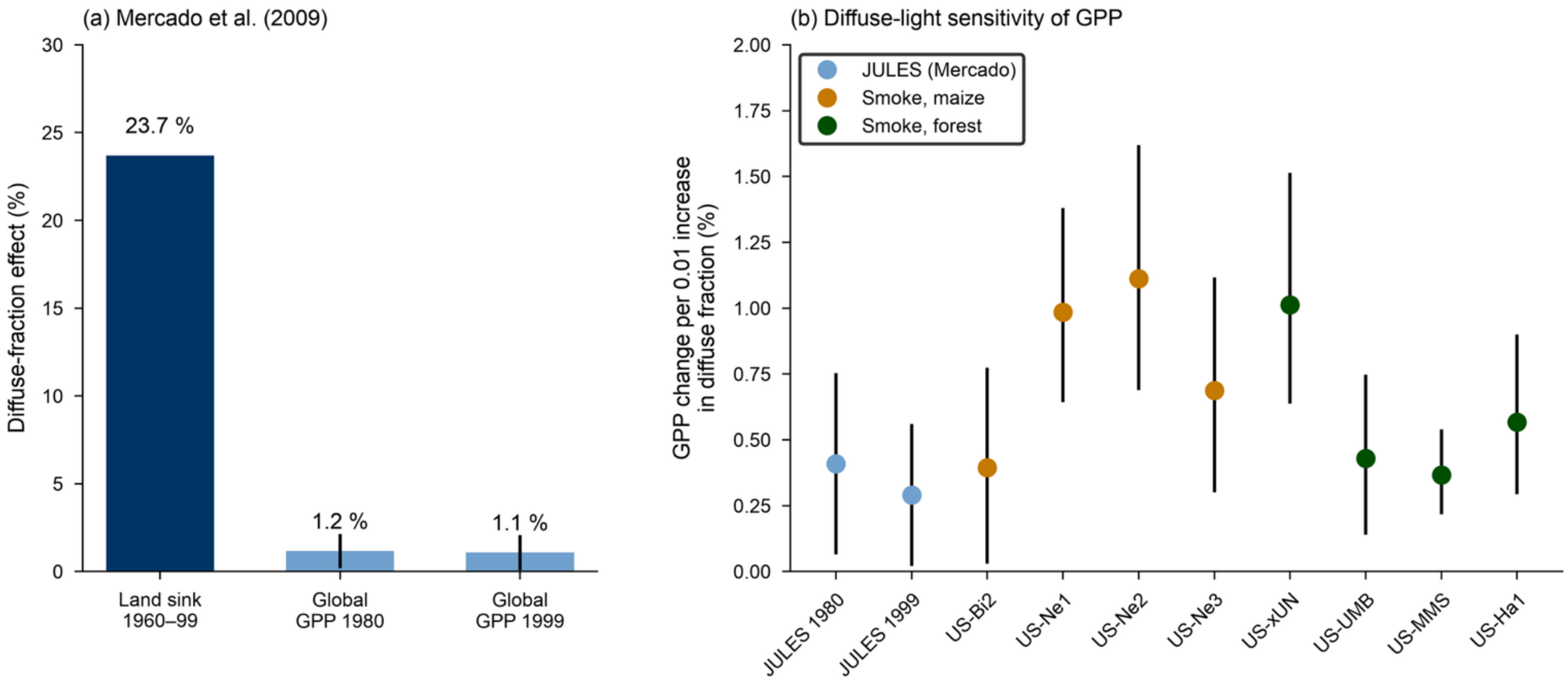

**Figure 6. Small GPP changes from diffuse light can drive large changes in the land carbon sink.**

(a) The change in land carbon sink from diffuse radiation from years 1960 to 1999 reported by Mercado et al.[9] and the corresponding change in GPP that led to that increase, derived from Mercado et al. Figure S10. (b) The normalized change in GPP per unit increase of diffuse fraction for Mercado et al.[9] 1980 and 1999, our regression, and Gu et al.'s default regression. For Mercado et al. this is expressed as a change in global GPP, with a baseline GPP of 123 PgC /yr[14]. For our regression, it is calculated at each site as the diffuse-fraction effect on GPP divided by the mean difference in diffuse fraction between smoke time steps and their matched clear-sky references. Error bars are the 95 % confidence intervals for our sites, and the range of the Figure S10 color classes for Mercado et al.[9].

## Discussion

Here, we use wildfire smoke as a natural analog for SAI, demonstrating that its radiative effects are similar to those of sulfate aerosols, resulting in an increase of GPP by an average of 7.9% across maize sites and 7.5% across forest sites. Our analysis suggests radiative effects from SAI would range from neutral to slightly positive, with potentially large increases in the land carbon sink. Correcting limitations in previous studies brings modeling and observational evidence into closer agreement. Together, wildfire smoke, volcanic eruptions, and air pollution all indicate that aerosol radiative effects are generally positive for plants.

While our findings utilize all available AmeriFlux data from sites that include diffuse radiation sensors, we are still limited by the number of smoke events that occur over these sites. AERONET instruments are not always near a tower site, so some of these measurements constitute a regional estimate (Extended Data Table 1). The scattering properties of smoke particles depend on the age of smoke as it undergoes physical and chemical evolution in the atmosphere, which could also impact results. Smoke can also impact pollutants, such as surface ozone and leaf deposition, which we do not account for, but this would make our estimates conservative, and previous studies have found ozone impacts from smoke to have a relatively small impact on photosynthesis[15].

Our results focus on temperate plant species in the United States, but limited observational data are available to assess whether these effects extend to tropical and boreal ecosystems. However, evidence from biomass-burning smoke suggests that tropical species also exhibit strong positive photosynthetic responses[16]. Smoke, volcanic eruptions, and air pollution tend to be short exposure events, and there is so far no research addressing how a sustained radiative perturbation, like that of SAI, lasting years or decades might affect plants. Long term laboratory and field studies (such as shading experiments) are needed to understand potential long-term effects of SAI. Unlike wildfire smoke, SAI would also alter precipitation patterns, introducing additional uncertainty due to differences and limitations among climate model projections. SAI would also alter the spectrum of light reaching the surface, and so far no work has been done to understand how that may affect plant species.

## Methods

### Flux tower data

This work utilizes eddy covariance flux tower data at eight different sites: US-Bi2, US-xUN, US-UMB, US-Ne1, US-Ne2, US-Ne3, US-MMS, and US-Ha1. We obtained the following variables from FLUXNET: GPP (GPP_NT_VUT_REF), PAR (PPFD_IN), diffuse fraction (DF, defined as: PPFD_DIF / PPFD_IN), shortwave radiation (SW_IN), top of atmosphere shortwave radiation (SW_IN_POT), temperature (TA_F_MDS, °C), vapor pressure deficit (VPD_F_MDS, hPa), and soil water content (SWC_F_MDS_1, %). Some sites report data in half-hourly timesteps (US-Bi2, US-xUN, US-UMB), while others report it in hourly timesteps (US-Ne1, US-Ne2, US-Ne3, US-MMS, and US-Ha1). A list of all sites and features can be found in Extended Data Table 1.

### Cloud free and smoke identification

For the analysis isolating the effects of smoke, we focus on cloud-free days to not confound the influences of smoke and clouds on GPP. Cloud-free days are days when the following criteria are met: SW_IN_POT > 50 W $m^{-2}$, Clearness index, kt = SW_IN / SW_IN_POT ≥ 0.45, kt changes by less than 0.03 (half-hourly) or 0.04 (hourly) compared with both the previous and the next time step to indicate a steady sky condition, at least one AERONET observation in the time step, and SW_IN must be at most 5 % above the highest clear-sky value normally seen at that time of day and time of year (the 98th percentile across all years, smoothed over ±10 days), because higher readings mean nearby clouds are reflecting extra light onto the sensor. In addition to clear sky conditions, partly cloudy time steps are also utilized to help fit the regression model used. Partly cloudy conditions are identified as kt ≥ 0.4.

Smoke events are classified by combining AOD measurements near sites and NOAA's Hazard Mapping System Fire and Smoke Product (HMS). A smoke event is identified on a clear sky day when the 550 nm AOD > 0.2, Ångström exponent > 1.0, and an HMS smoke flag is present on that day at a site. For each flux tower, AOD and Ångström exponent are the average of the measurements from its nearby AERONET stations (Table 1), with each station weighted equally. US-xUN and US-Ha1 have an AERONET station at the tower, so their values come from that single station. To compare to a clear sky reference for each smoke time step, we use all clean reference (AOD < 0.15) steps at the same site within ±15 days of the date and within ±1 hour of the time of day. At least 3 measurements are required, otherwise the smoke time step is dropped. Reference values for GPP and of every driver are the means over these references.

To fully understand the effects of radiation changes from smoke, a filter is applied to the data before analysis. Since some crop sites rotate between crops, we keep only years where the USDA Cropland Data Layer gives corn at a tower. To ensure that there is something growing at a site we use PhenoCam measurements at each tower which photographs the canopy every day and records how green it is. We keep only the days when the canopy is at least 30 % of the way from that year's least green to its most green, so bare fields and leafless forest are left out. To isolate the GPP responses due to changes in incident radiation, we filter to remove periods of temperature stress (outside 20-30 °C) or high atmospheric demand (VPD > 3 kPa).

**Modeling the impact of smoke on GPP**

The regression model used to understand the effects of smoke on GPP through changes in different drivers is an ordinary least squares model fitted to each site (Eq. 1), where D is the PAR diffuse fraction, VPD is vapor pressure deficit (hPa), TA is surface temperature (°C), and SWC is soil water content (%), $b_0$–$b_8$ are coefficients fitted separately at each site, and ε is the residual.

$$\mathrm{GPP} = b_0 + b_1 P + b_2 P^2 + b_3 \mathrm{D} + b_4 P\ \mathrm{D} + b_5 \cos\theta + b_6 \mathrm{VPD} + b_7 \mathrm{TA} + b_8 \mathrm{SWC} + \varepsilon \qquad [1]$$

$$\text{with } P = \frac{\mathrm{PAR}}{1000} (mmol\ m^{-2} s^{-1}) \text{ and } \theta = \text{solar zenith angle}$$

The fitting sample includes all cloud-free time steps (at all aerosol levels) and partly cloudy time steps (kt ≥ 0.4). Partly cloudy time steps are included only if their diffuse fraction does not exceed the 98th percentile of diffuse fraction during smoke, so the model is trained across the range of light conditions found under smoke without extrapolating to overcast skies.

To attribute the change in GPP from smoke to the individual drivers, we match each value from the smoke time step, s, to the mean values of its clear-sky references, r. Let f(P, D) be the GPP predicted by the site regression (Eq. 1) with PAR = P and diffuse fraction = D, and all other drivers at their smoke values. The change in GPP from the change in light was split into a contribution from less total light and a contribution from more diffuse light. Because the effect of total light depends on the diffuse fraction and vice versa, each contribution was calculated twice, with the other variable held at its clear-sky value and at its smoke value, and the two results were averaged; the total light (Eq. 2) and diffuse light (Eq. 3) contributions then sum exactly to the total radiation effect (Eq. 4).

$$\Delta \mathrm{GPP}_{\mathrm{light}} = \frac{1}{2}[f(P_s, D_s) - f(P_r, D_s) + f(P_s, D_r) - f(P_r, D_r)] \qquad [2]$$

$$\Delta \text{GPP}_{\text{diffuse}} = \frac{1}{2}[f(P_s, D_s) - f(P_s, D_r) + f(P_r, D_s) - f(P_r, D_r)] \quad [3]$$

$$\Delta \text{GPP}_{\text{rad}} = \Delta \text{GPP}_{\text{light}} + \Delta \text{GPP}_{\text{diffuse}} = f(P_s, D_s) - f(P_r, D_r) \quad [4]$$

Where subscripts s and r denote values at the smoke time step and the mean of its clear-sky reference steps, respectively. The effects of VPD, air temperature and soil moisture were estimated one driver at a time. For each driver X (VPD, TA, SWC), X was changed between its smoke and reference value while all other drivers were held at their smoke values, and again while they were held at their reference values. The two changes were averaged:

$$\Delta \text{GPP}_X = \frac{1}{2}[f(x_s) - f(x_s | X = X_r) + f(x_r | X = X_s) - f(x_r)] \quad [5]$$

where $x_s$ and $x_r$ are the full sets of smoke and reference driver values. Each effect k is expressed as a percentage of reference GPP, summed over all n matched smoke steps i:

$$\Delta \text{GPP}_k(\%) = 100 \times \frac{\sum_{i=1}^{n} \Delta \text{GPP}_{k,i}}{\sum_{i=1}^{n} \text{GPP}_{r,i}} \quad [6]$$

The total GPP response is then the sum of the radiation, VPD, temperature and soil-moisture effects.

**Radiative transfer modeling to compare smoke with sulfate**

We used the libRadtran radiative transfer model (uvspec, version 2.0.6)[10,17] with the DISORT solver[18] to quantify how additional aerosols alter the amount and composition of sunlight reaching the surface. We compared wildfire smoke with stratospheric sulfate aerosols, the proposed aerosol for SAI. For each aerosol type, we calculated direct and diffuse downward irradiance at the surface across the photosynthetically active radiation (PAR; 400–700 nm). Simulations covered added aerosol optical depths (AOD at 550 nm) of 0–2 and solar zenith angles of 20–60°. Both aerosol types were added to an identical clean-sky background aerosol. Background and smoke optical properties were derived from the eight study sites using median AERONET observations across all clean-reference and smoke time steps, respectively.

Stratospheric sulfate was represented by a lognormal size distribution with an effective radius of 0.4 μm and a geometric standard deviation of 1.5, characteristic of the aerosol produced by the 1991 Mt. Pinatubo eruption, and a refractive index of 1.43. Its optical properties were calculated using Mie theory[19], yielding an asymmetry parameter of 0.75 and an Ångström exponent of 0.50. Sulfate was treated as non-absorbing, with a single-scattering albedo (SSA) of approximately 1. An effective radius was found to produce the most favorable effects on diffuse fraction for loss of PAR, so the size of sulfate would partly determine vegetative impacts (Extended Data Figure 4). The radiation effect on GPP became negative only when strong absorption was combined with high AOD, conditions beyond those observed for our smoke events (Extended Data Figure 5). Smoke optical properties were taken as the median AERONET retrievals across all smoke time steps at the eight sites (SSA 0.96, asymmetry parameter 0.65, Ångström exponent 1.69). Both aerosols were added to a clean background aerosol, defined as the median AERONET retrieval at clean reference time steps (AOD 0.076, SSA 0.94, asymmetry parameter 0.67, Ångström exponent 1.21). The combined SSA and Ångström exponent were weighted by optical depth, and the asymmetry parameter by scattering optical depth. For each aerosol type, we calculated two quantities across the range of added aerosol optical depths (AOD, 0–2 at 550 nm) and solar zenith angles (20–60°). The first quantifies the fraction of clean-sky irradiance reaching the surface:

$$\mathrm{frac}(x) = \frac{E_{\mathrm{dir}(x)} + E_{\mathrm{dif}(x)}}{E_{\mathrm{dir}(0)} + E_{\mathrm{dif}(0)}} \qquad [7]$$

The second is the change in diffuse fraction:

$$\Delta\mathrm{DF}(x) = \frac{E_{\mathrm{dif}(x)}}{E_{\mathrm{dir}(x)} + E_{\mathrm{dif}(x)}} - \frac{E_{\mathrm{dif}(0)}}{E_{\mathrm{dir}(0)} + E_{\mathrm{dif}(0)}} \qquad [8]$$

Here E_dir and E_dif are the direct and diffuse irradiance and x is the added AOD. Applying these changes to the clear-sky reference conditions at each site, through the site regressions, gives the modelled GPP response to smoke and sulfate seen in Fig. 3.

**Comparing to Gu et al.**

For the analysis comparing our results with Gu et al.[2], we did not apply the weather filters (temperature of 20–30 °C and VPD > 3 kPa) used in the preceding analyses to the Harvard Forest data. These filters were not applied in Gu et al.[2], and we therefore omitted them to maintain consistency with their methodology. We first use the default Gu et al.[2] model (SOM Eq. 2) with their default parameters (Gu et al.[2], Table 1, 1992) to understand how smoke at Harvard Forest would impact GPP through changes in radiation compared to our model (Eq. 1). Next, we kept the Gu et al.[2] model (SOM Eq. 2) unchanged and refit its parameters from Gu et al.[2] Table 1 using the Harvard Forest clear sky data from 2012-2025 (Extended Data Table 2). Smoke increases GPP in this model by turning direct light into diffuse light, so the size of the gain depends on how much better the canopy uses diffuse light than direct light. In Gu et al.'s 1992 parameters, the diffuse CLR is seven times the direct CLR (where CLR indicates how much canopy photosynthesis can still rise in bright light before it saturates). Refit to the Harvard Forest tower, it is only about twice the direct value (Extended Data Table 2). Moving light from direct to diffuse therefore adds less GPP, and the smoke effect falls from +23 % to +13 % (using clear sky data). Partly cloudy time steps cover a much wider range of diffuse fractions than clear skies, including those seen during smoke, which lets the fit separate the diffuse and direct parameters. Including partly cloudy conditions to fit the parameters causes the radiative smoke effect to drop again from +13% to +9% (Figure 4a).

The light response model of Gu et al.[2] (SOM eq. 2) has only light inputs. When diffuse light rises, temperature and VPD can fall at the same time, and these changes also affect GPP. With no weather terms, the fit must attribute all of that GPP change to light. To test if weather was confounding its results, we also fitted the model with added linear terms for VPD, air temperature (T_a), soil moisture (SWC) and solar zenith angle (θ), using all clear and partly cloudy time steps:

$$\mathrm{GPP} = P\left(I_f, I_r\right) + c_0 + c_1 \cos\theta + c_2 \mathrm{VPD} + c_3 \mathrm{TA} + c_4 \mathrm{SWC} \qquad [9]$$

**Revisiting Proctor et al.**

To understand why Proctor et al.[1] find a large insolation loss for maize, we used their replication data (FAO national maize yields, 1980–2005) and their yield model (SI eq. 16). The model regresses log yield on stratospheric aerosol optical depth (SAOD). It controls for growing-season temperature, precipitation and cloud fraction (restricted cubic splines) and for ENSO, with country fixed effects and country-specific quadratic time trends. Standard errors are clustered by country and by year. This reproduced their maize SAOD coefficient exactly (−0.649). We re-estimated SI eq. 16 of Proctor et al.[1] unchanged but weighted each country-year by the country's mean maize harvested area (1980–2005) using Eq. 10, so that countries count in proportion to how much maize they grow. Proctor et al. weighted all countries equally.

$$\min_{\beta,f,g,\lambda,\phi} \sum_{i,t} \bar{\mathrm{A}}_i \epsilon_{it}^2 \qquad [10]$$

where $\bar{A}_i$ is the mean maize harvested area of country i over 1980–2005 (FAO), the sum is over all countries i and years t, and the coefficients of Proctor et al.'s Eq. 16 (β, f_T, f_P, f_C, g, $\lambda_i$, $\phi_{i1}$ and $\phi_{i2}$) are chosen to minimize the area-weighted sum of squared residuals $\epsilon_{it}$. This weighting gives the effect of stratospheric aerosols on maize yield per hectare of maize grown worldwide, rather than the average effect across countries, which makes the result more comparable with crop modelling studies that report area-weighted global yield changes.

To identify the yield anomaly missed by the controls from Proctor et al.'s[1] Eq. 16, we removed the $\beta S_{it}$ term. The leftover $\epsilon_{it}$ for each country and year is the yield change the controls don't explain. Averaging it by year showed a large 1992 loss in southern Africa that the controls miss. We then kept their Eq. 16 the same, but removed the nine southern Africa countries (South Africa, Zimbabwe, Zambia, Malawi, Mozambique, Botswana, Namibia, Lesotho and Eswatini) from the data. To check that removing southern Africa isn't just a random effect of removing any nine countries, Eq. 16 in Proctor et al.[1] was refit 10,000 times. Each time, nine randomly chosen countries were dropped instead:

$$\ln(Y_{it}) = f_T(T_{it}) + f_P(P_{it}) + f_C(C_{it}) + g(E_t, E_{t-1}, i) + \beta S_{it} + \lambda_i + \phi_{i1} t + \phi_{i2} t^2 + \epsilon_{it}, \quad i \notin D_m \qquad [11]$$

where $D_m$ is the random set of nine countries dropped in draw $m$. The p-value is the share of draws that raised β at least as much as dropping southern Africa did:

$$p = \frac{1 + \#\{\Delta\beta_m \geq \Delta\beta_{SA}\}}{N + 1} \qquad [12]$$

where N = 10,000 is the number of draws, $\Delta\beta_m$ is the change in β in draw $m$, $\Delta\beta_{SA}$ is the change in β when southern Africa is dropped, and #{ } counts the draws that meet the condition. None of the 10,000 random draws changed β as much as dropping southern Africa, so p = 1/10,001 ≈ 0.0001 (Extended Data Figure 3).

**Estimating potential effects of stratospheric aerosol injection**

For understanding the potential effects of stratospheric aerosol injection on plant growth using our relationships built from AmeriFlux observations, we used output from the Community Earth System Model version 2 (CESM2), Whole Atmosphere Community Climate Model version 6 (WACCM6). CESM2 simulated the Geoengineering Modeling Intercomparison Project scenario G6-1.5K-SAI, which followed the medium emissions pathway SSP2-4.5. G6-1.5K-SAI injects $SO_2$ in equal amounts at 30°N and 30°S to maintain global average temperatures of 1.5K above preindustrial. More information on the scenario design and simulations can be found in Lee et al.[13]. We used daily output of surface shortwave and diffuse radiation at grid cells corresponding to each AmeriFlux site and averaged them by month over 2065–2084 for the SAI and SSP2-4.5 runs. For each site and calendar month, the change in light is the ratio of SAI to SSP2-4.5 shortwave, and the change in diffuse fraction is their difference in diffuse ÷ shortwave. These changes were applied to the tower PAR and diffuse fraction for that month, and the site regression gave the resulting change in GPP. An overview of changes in light seen at each tower site under SAI can be found in Extended Data Table 3.

## Data availability

All of the processed data to replicate the findings here is available at https://github.com/bjc204/SAI_analogs. FLUXNET-format eddy covariance data for US-Bi2, US-Ne1, US-Ne2, US-Ne3, US-xUN, US-UMB, US-MMS and US-Ha1 are publicly available from the AmeriFlux data portal (https://ameriflux.lbl.gov). AERONET aerosol optical depth and inversion products are available from NASA AERONET (https://aeronet.gsfc.nasa.gov). NOAA Hazard Mapping System smoke polygons are available from NOAA NESDIS (https://satepsanone.nesdis.noaa.gov/pub/FIRE/web/HMS/). Canopy greenness data are from the PhenoCam Network (https://phenocam.nau.edu). The Proctor et al.[1] replication data are available from Zenodo (https://doi.org/10.5281/zenodo.1313800). CESM2-WACCM6 G6-1.5K-SAI and SSP2-4.5 simulations were obtained from Lee et al.[13] (https://zenodo.org/records/17613419).

## Code availability

All the code to replicate the findings here is available at https://github.com/bjc204/SAI_analogs.

## Acknowledgments

BC and DV were supported by the Quadrature Charitable Foundation. BK, DV, and ML were supported by the Environmental Defense Fund. ML was also supported by the Morven Sustainability Lab of the University of Virginia.  Claude Opus 5.5 was used to assist with code development and figure generation. The authors take full responsibility for all code and results presented here. We thank the AmeriFlux community and the individuals who maintain and support the AmeriFlux network and data infrastructure, which helped make this research possible.

## Contributions

BC conceptualized the idea and designed the study, with input from BK. BC performed the research and created visualizations. BK, DV, and ML acquired funding. BC wrote the first draft, and DV, BK, and ML reviewed and edited the paper.

## Competing Interests

The authors declare no competing interests.

## Corresponding Author

Correspondence to Brendan Clark (b.clark.ma@gmail.com).

## Extended Data for "Natural analogs for geoengineering suggest modest radiative plant gains"

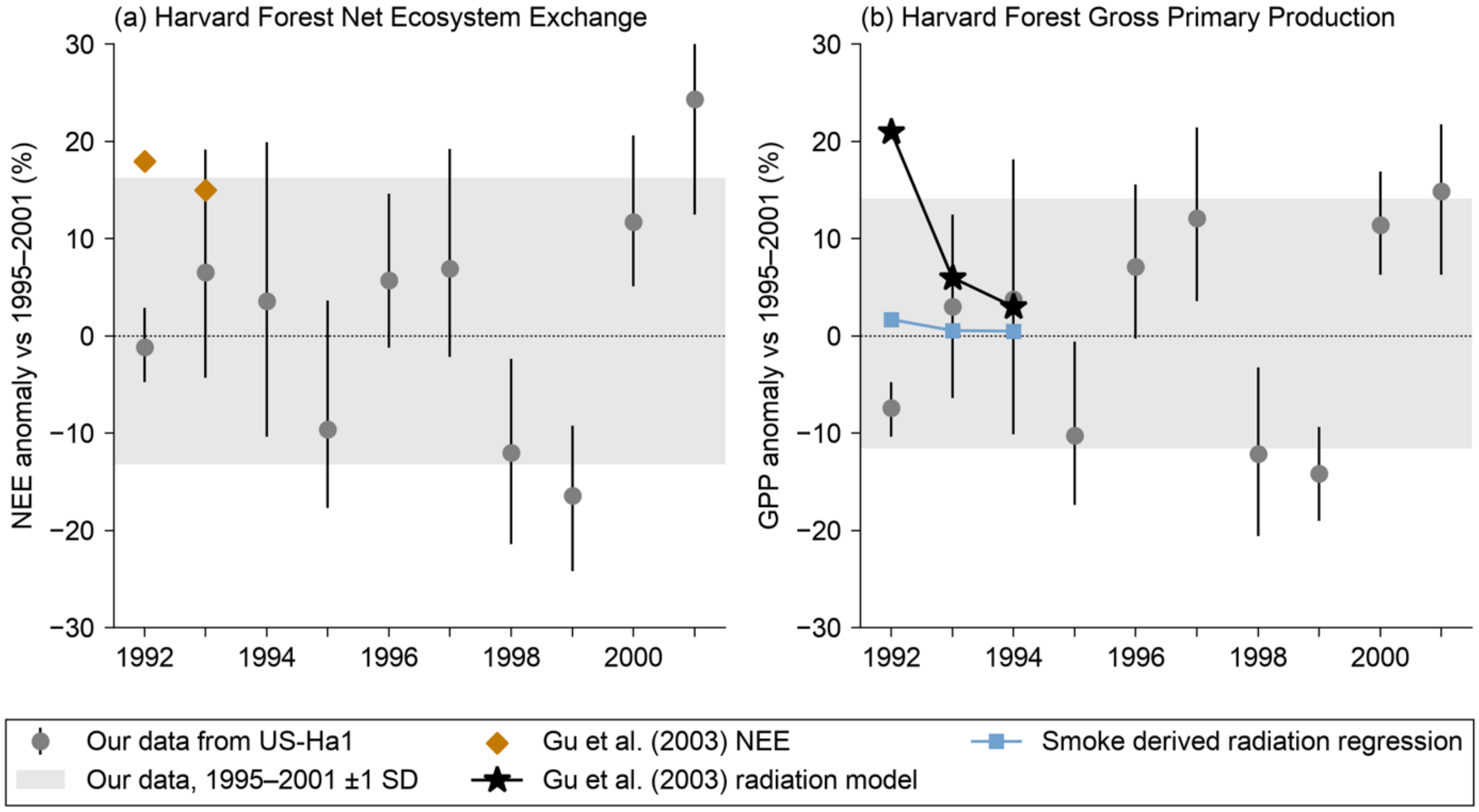

**Extended Data Figure 1. No detectable increase in Harvard Forest carbon uptake after Pinatubo in current flux data.**

Clear-sky anomalies in (a) net ecosystem exchange (NEE) and (b) gross primary production (GPP) at Harvard Forest (US-Ha1) from 1992–2001, relative to the 1995–2001 mean. Grey points in (a) show NEE uptake anomalies calculated from current FLUXNET data using the methodology of Gu et al. (2003), with 95% confidence intervals estimated using a day-block bootstrap. The grey shaded band represents ±1 standard deviation of the 1995–2001 anomalies. Blue squares in (b) show GPP anomalies predicted by our smoke-derived regression model, applied to the observed increase in diffuse radiation in each year. For comparison, orange diamonds in (a) show the NEE increases reported by Gu et al. (2003), while black stars in (b) show their modelled changes in canopy photosynthesis.

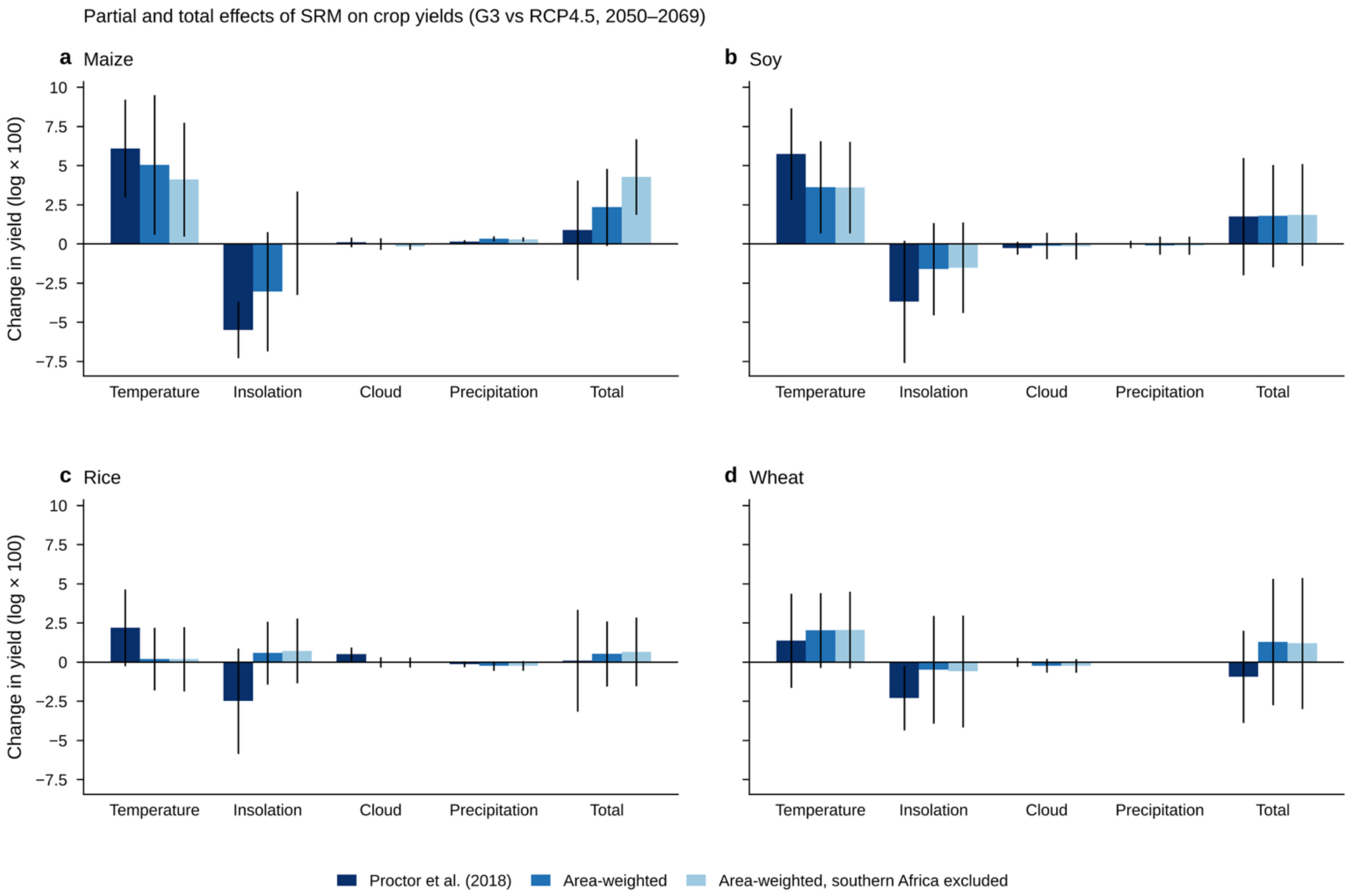

**Extended Data Figure 2. Area weighting removes the insolation losses for all four crops.**

Same as Figure 4c, but for all crops in Proctor et al. (2018)'s analysis.

**Extended Data Table 1 | Flux tower sites, vegetation and aerosol stations**

| Site | Location | Land cover | Vegetation | Canopy height (m) | Peak LAI ($m^2$ $m^{-2}$) | AERONET sites (distance, km) |
|---|---|---|---|---|---|---|
| US-Bi2 | Bouldin Island, California | Cropland | Maize, irrigated | 3.2 | 2.7 | Sacramento_River (32), Grizzly_Bay (46), Sacramento_CARB (51), Modesto (70), Turlock_CA_USA (88), NASA_Ames (89) |
| US-Ne1 | Mead, Nebraska | Cropland | Maize, irrigated | 3.5 | 5.3 | NEON_KONZ (230), Ames (244), NEON_UKFS (260), Sioux_Falls (286) |
| US-Ne2 | Mead, Nebraska | Cropland | Maize, irrigated | 3.4 | 4.9 | NEON_KONZ (230), Ames (244), NEON_UKFS (260), Sioux_Falls (286) |
| US-Ne3 | Mead, Nebraska | Cropland | Maize, rainfed | 3.0 | 4.5 | NEON_KONZ (231), Ames (241), NEON_UKFS (260), Sioux_Falls (285) |
| US-xUN | UNDERC, Wisconsin–Michigan | Mixed forest | Red maple, sugar maple, aspen, paper birch; balsam fir | 24.0 | 6.4 | NEON_UNDE (<1) |
| US-UMB | UMBS, Michigan | Deciduous broadleaf forest | Bigtooth and trembling aspen with northern hardwoods | 20 | 6.7 | Granite_Island (245), South_Greenbay (276) |
| US-MMS | Morgan-Monroe State Forest, Indiana | Deciduous broadleaf forest | Secondary hardwoods, maple–beech to oak–hickory (29 tree species) | 27 | 6.4 | BONDVILLE (186), Dayton (204) |
| US-Ha1 | Harvard Forest, Massachusetts | Deciduous broadleaf forest | Red oak, red maple; patches of hemlock and white pine | 25.4 | 6.5 | NEON_Harvard (<1) |

**Extended Data Table 1.** A list of all AmeriFlux sites used in the analysis, a description of the site, and a list of all nearby AERONET sites used for AOD measurements.

**Extended Data Table 2 | Parameters of the Gu et al. (2003) canopy model at Harvard Forest**

| Parameter | Gu et al. (2003), 1992 | Refit, clear sky | Refit, clear and partly cloudy |
|---|---|---|---|
| Initial diffuse RUE (μmol $J^{-1}$) × 100 | 11.82 | 10.30 | 8.20 |
| Initial direct RUE (μmol $J^{-1}$) × 100 | 3.94 | 3.27 | 5.91 |
| Diffuse CLR (μmol $m^{-2}$ $s^{-1}$) | 139.60 | 64.63 | 70.69 |
| Direct CLR (μmol $m^{-2}$ $s^{-1}$) | 19.33 | 31.24 | 26.44 |

**Extended Data Table 2.** The parameters used in the Gu et al. (2003) model by default (left), by refitting the parameters to 14 years of growing season, clear sky data (middle) and 14 years of clear sky and partly cloudy data (right).

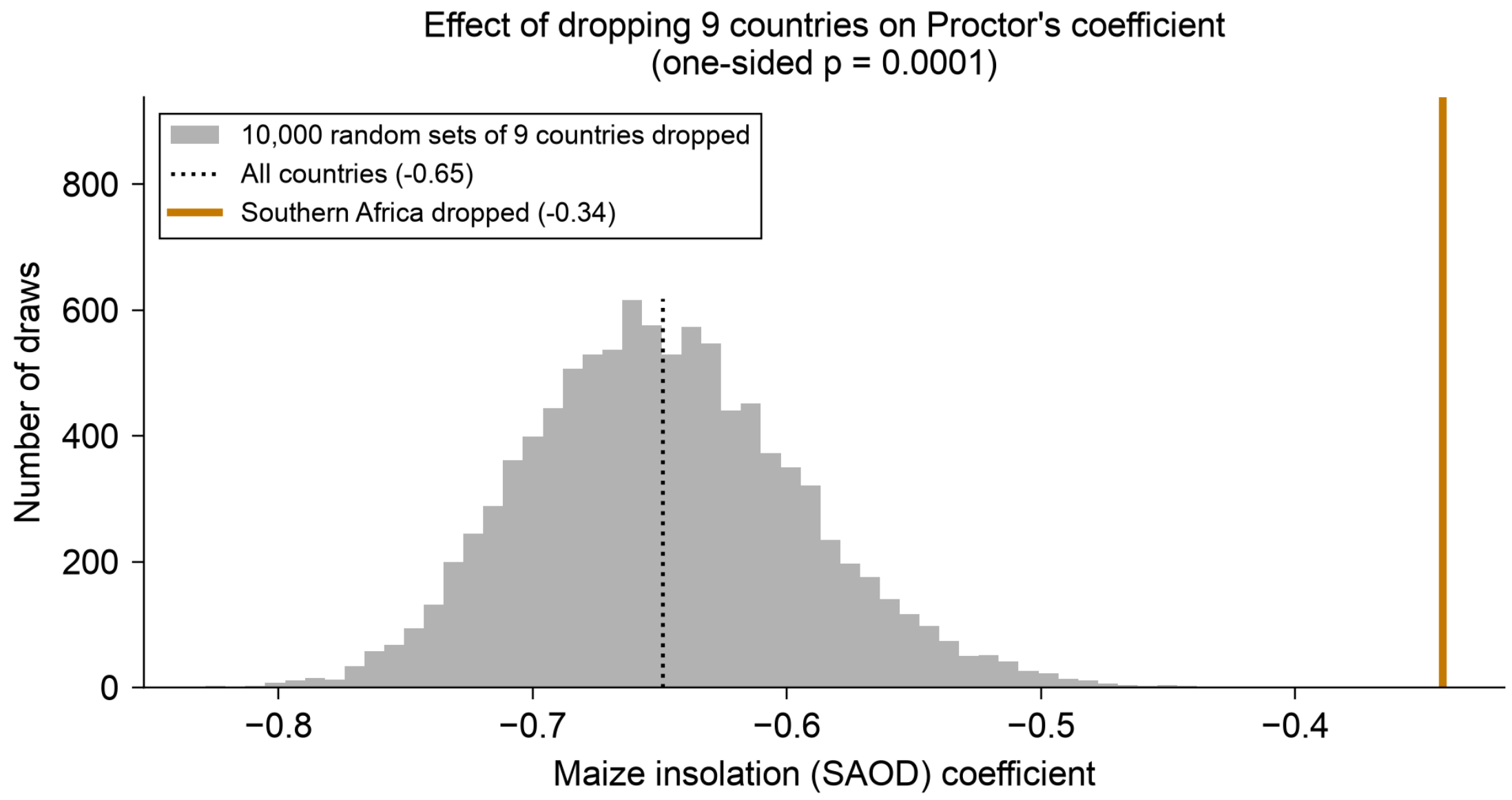

**Extended Data Figure 3. Excluding southern Africa halves the maize insolation effect, far more than excluding other countries.**

The effect on the insolation (SAOD) coefficient used in Proctor et al.'s non-area weighted regression model from randomly dropping 9 countries 10,000 times, compared with dropping the 9 countries in Southern Africa (orange line), using the default non-area weighting.

**Extended Data Table 3 | CESM2 surface shortwave and diffuse radiation at each site, SSP2-4.5 and SAI (2065–2084)**

| Site | Months | Shortwave (W $m^{-2}$) | | Diffuse (W $m^{-2}$) | | Diffuse fraction | | Change, SAI vs SSP2-4.5 | |
|---|---|---|---|---|---|---|---|---|---|
| | | SSP2-4.5 | SAI | SSP2-4.5 | SAI | SSP2-4.5 | SAI | Light (%) | Diffuse fraction (%) |
| US-Bi2 | Mar–Sep | 291 | 286 | 50 | 68 | 0.18 | 0.25 | −1.5 | +39 |
| US-Ne1 | May–Sep | 264 | 258 | 70 | 83 | 0.26 | 0.32 | −2.2 | +23 |
| US-Ne2 | Jun–Sep | 264 | 254 | 66 | 81 | 0.25 | 0.32 | −3.6 | +26 |
| US-Ne3 | Jun–Sep | 264 | 254 | 66 | 81 | 0.25 | 0.32 | −3.6 | +26 |
| US-xUN | May–Sep | 248 | 241 | 69 | 83 | 0.28 | 0.34 | −2.8 | +23 |
| US-UMB | May–Oct | 229 | 223 | 65 | 78 | 0.29 | 0.36 | −2.6 | +24 |
| US-MMS | Mar–Nov | 218 | 212 | 62 | 75 | 0.28 | 0.35 | −2.4 | +25 |
| US-Ha1 | May–Oct | 219 | 218 | 69 | 80 | 0.31 | 0.36 | −0.9 | +18 |

**Extended Data Table 3.** Changes in light under G6-1.5K-SAI minus SSP2-4.5 (2065-2084 average) from one ensemble member of CESM2-WACCM6 (Lee et al., 2026). Months are the growing season of each respective site.

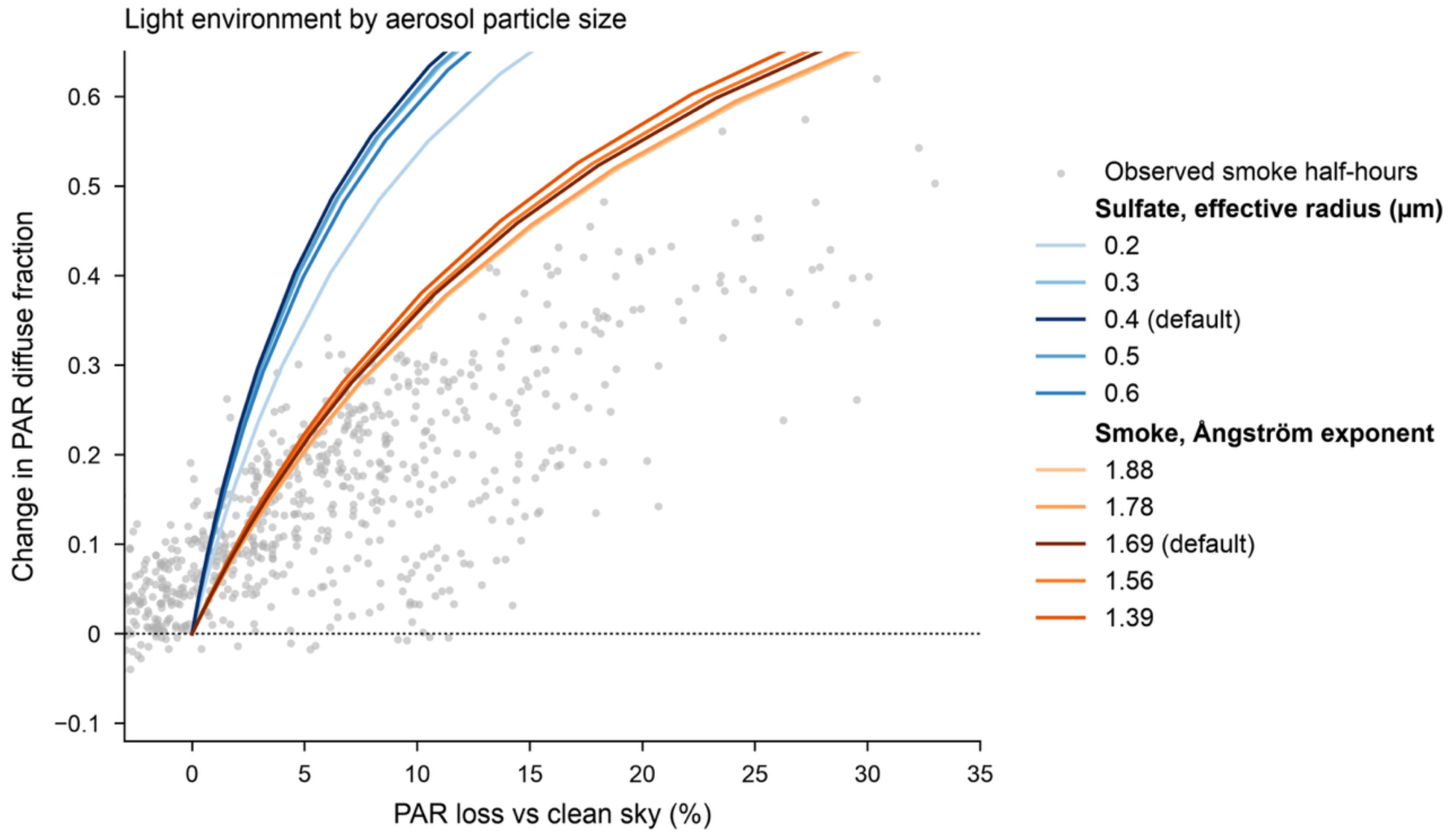

**Extended Data Figure 4. Sensitivity of the light environment to aerosol particle size.**

Modeled change in PAR diffuse fraction against PAR loss relative to clean sky as aerosol optical depth increases. Blue lines show sulfate aerosol with effective radii of 0.2–0.6 μm (default 0.4 μm). Orange lines show smoke with Ångström exponents of 1.39–1.88 (default 1.69). Gray points are observed smoke-affected half-hours at the AmeriFlux sites.

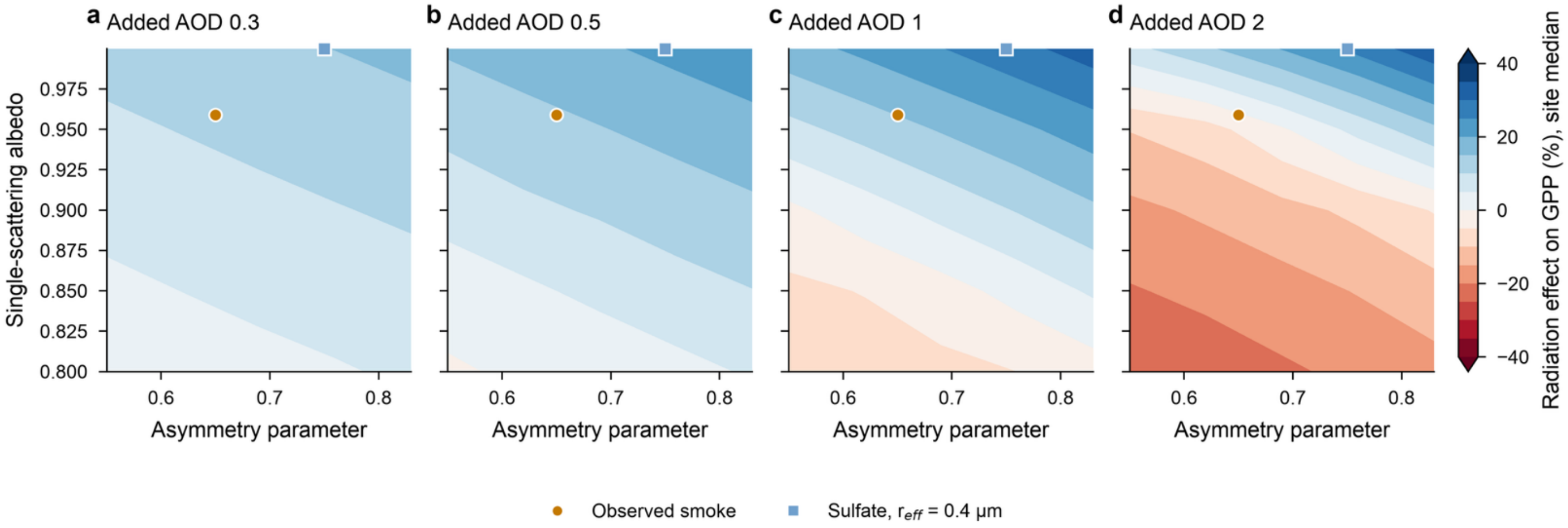

**Extended Data Figure 5. The radiation effect on GPP turns negative only for strongly absorbing aerosol or very high aerosol loading.**

Panels a–d show the site-median radiation effect on GPP (%) across asymmetry parameter and single-scattering albedo for added aerosol optical depths of (a) 0.3, (b) 0.5, (c) 1, and (d) 2. Blue shading marks GPP gains and red shading marks losses. Circles mark the median single-scattering albedo (0.96) and asymmetry parameter (0.65) of observed smoke, taken across all smoke events at the study sites from AERONET retrievals. Squares mark sulfate with an effective radius of 0.4 μm, as simulated in libRadtran. Aerosols that scatter more and scatter more strongly forward (upper right) give larger GPP gains. At low to moderate AOD, both aerosol types increase GPP. At AOD 2, the smoke effect falls to about zero, while sulfate still raises GPP.